# Learning-based real-time method to looking through scattering medium beyond the memory effect

Enlai Guo, Shuo Zhu, Yan Sun, Lianfa Bai, Jing Han*

Jiangsu Key Laboratory of Spectral Imaging and Intelligent Sense, Nanjing University of Science and Technology, Nanjing 210094, China

*: Corresponding author :eohj@njust.edu.cn;

**Abstract:** Strong scattering medium brings great difficulties to optical imaging, which is also a problem in medical imaging and many other fields. Optical memory effect makes it possible to image through strong random scattering medium. However, this method also has the limitation of limited angle field-of-view (FOV), which prevents it from being applied in practice. In this paper, a kind of practical convolutional neural network called PDSNet is proposed, which effectively breaks through the limitation of optical memory effect on FOV. Experiments is conducted to prove that the scattered pattern can be reconstructed accurately in real-time by PDSNet, and it is widely applicable to retrieve complex objects of random scales and different scattering media.

## 1 Introduction

Scattering medium exists generally in biological tissues and is the main interference source in the field of astronomical imaging. Many new imaging methods have been proposed and improved to realize imaging through heavily disordered medium. The typical ones include optical coherence tomography, wavefront modulation, image reconstruction based on transmission matrix, and reconstruction based on point spread function (PSF). Based on optical memory effect (OME), Katz proposed speckle correlation imaging technology[1,2]. Different from the above methods, this method provides powerful capability to image through scattering media deeply, can be applied to dynamic scattering media, and no additional reference source is needed.

Due to the existence of OME, scattering medium can be regarded as a linear system within a settled FOV. Based on this principle, two kinds of algorithms have emerged: the first kind is based on the deconvolution of the PSF, which requires the measurement of the system PSF in advance and belongs to the invasive method. The second type is speckle correlation imaging technology, which uses autocorrelation to concatenate speckle pattern with the original target directly, and finally utilizes phase recovery algorithm to reconstruct the target. At present, typical phase recovery algorithms include HIO[3], prGAMP[4], BM3D-prGAMP[5], etc. The FOV of an imaging system is strictly limited by OME and is inversely proportional to the effective thickness of the scattering medium, which limits the anti-scattering ability of various OME-based algorithms to a certain extent. The FOV of OME has been successfully expanded through several novel techniques, but most of them belong to invasive method, which rely on priori knowledge[6-8].As far as we know, the best non-invasive method to imaging beyond the OME range, is a double loop iterative algorithm proposed by Dai[9], which can restore two targets respectively whose total size is beyond OME, the algorithm can expand at least three times of OME scope shown by experiments, no prior information is required, but there are still some constraints for the target to be restored : the target must be in regular shape and in two independent OME region. Since double iterative optimization is required, the algorithm is time-consuming, it needs 14400s to reconstruct the image with resolution of 300*300 pixels on MATLAB2018b.

To solve the original target distribution from the scattering image is to find the mapping relationship between them. Many scholars have also applied the deep learning algorithm to image through scattering medium in the context of OME, to solve the distribution of tested targets. "IDiffNet" proposed by Barbastathis et al. realizes the reconstruction of speckle image [10], and discusses in detail the influence of loss function, training set and other variables on the final reconstruction image quality. Tian et al. constructed a CNN network [11] with the "one-to-all" reconstruction capability of single optical statistical characteristics, which can reconstruct the speckle image generated by untrained ground glass, only condition is that untrained scattering medium has the same statistical characteristics as the ground glass used in making the training set. The scattered images of ground glass and single mode fiber are recovered in utilizing U-net realized by Guo et al[12]. A GAN network is built by Zeng and fat emulsion solution is used to simulate dynamic scattering media[13]. The experiment proves that the network can realize the restoration of dynamic scattering images. Guohai Situ et al. constructs a hybrid neural network (HNN) model[14], which is proved that targets larger than the FOV of OME can also be reconstructed in using the HNN. The experiments in literature [14] all employ a single ground glass as scattering medium and the reconstruction ability of optical properties of different medium with one training process is not been proved to possess. The above works prove that deep learning can be used to reconstruct objects within the OME range through random scattering media and break through the OME range of a single scattering medium.

In order to introduce OME-based methods into living biological tissues and other practical applications, the limitation of this kind of method in FOV is a problem demanding prompt solution. At the same time, the structural complexity of the actual target is very high, so reconstruction ability is demanded to be powerful enough to retrieve the image of hidden objects. In practical application, the scattering medium is complex, both the scale of the target to be measured and the properties of the scattering medium are complex and changeable. Good generalization ability is also necessary for the algorithm. Therefore, breaking through the limitation of OME on the FOV, good reconstruction ability for complex targets and strong generalization ability, are the basic conditions for imaging method through scattering media to be practical.

In this paper, a kind of convolutional neural network named PDSNet (Pragmatic De-scatter ConvNet) is constructed to break through the limitation of OME on FOV in a data-driven way, combining with the principle of traditional speckle correlation imaging algorithm to guide the design and optimization of the network. PDSNet is a neural network structure suitable for random scales and complex targets.

## 2 Method

### 2.1 Basic theory

To break through the constraint of OME on the FOV, solving the original target distribution from the scattering image is the inverse problem corresponding to the imaging process, which can be expressed as follows:

$$O = F^{-1}(M) \tag{1}$$

Where, $O$ represents the target to be measured, $M$ represents the output result obtained through the optical system, i.e., the scattering image, and $F$ represents the forward model of the imaging process.

The inverse problem is essentially an optimization problem, which can be expressed as:

$$O = \arg\min \|F(O) - M\|^2 + \lambda R(O) \tag{2}$$

Where, $R(O)$ is the regularization term. Starting with the speckle image, the original target to be restored is solved, which is the optimization problem represented by the above equation.

The light signal loaded with target information finally reaches the camera after being modulated by the scattering medium. The process can be expressed as follows:

$$I = \sum_{i=1}^{n}(O_i * S_i) \tag{3}$$

Where, n represents the target distributed in n different OME ranges, $O_i$ is the ith imaging target, and $S_i$ is the corresponding PSF.

According to the literature [2], the autocorrelation of PSF S is an impulse function. When the target is completely in the range of OME, autocorrelation is performed on the speckle image, as follows:

$$I \star I \approx O \star O \tag{4}$$

It can be inferred that, if the target is distributed in the corresponding range of n optical memories, the autocorrelation distribution of the speckle image can be expressed as:

$$I \star I \approx \sum_{i=1}^{n}(O_i \star O_i) \tag{5}$$

Formula 5 shows that even if the target to be restored crosses several OME ranges at the same time, there is a strong physical constraint between the target and the scattered image. Therefore, the optimization problem corresponding to formula 2 should have an optimal solution.

The traditional algorithms used in speckle correlation imaging have various limitations in practical use: for example, HIO[3] needs multiple iteration cycles. This type of algorithm is highly dependent on the random initial values, so its optimization results often converge to the local optimal solution and take a long time. Although improved algorithms such as prGAMP[4] and BM3D-prGAMP[5] can achieve better reconstruction effects than traditional iterative phase recovery algorithms, they have not been improved for the time-consuming defects. Moreover, the reconstruction effects of such algorithms depend to some extent on prior information such as target sparsity, which is not conducive to practical use.

A large target beyond the range of OME can be seen as consisting of several small targets that satisfy the OME constraint. According to the existing traditional algorithm to solve optimization problems, the solution complexity of the optimization problem corresponding to formula 5 increases by an order of magnitude with the increase of n. Therefore, it is bound to introduce various constraints or prior information to assist the solution of the optimization problem, which will ultimately be more detrimental to the implementation and application of the technology. In terms of optimization problem solving, deep learning method is better at using a large amount of data and mining the potential mapping relationship between independent variables and dependent variables compared with traditional algorithms, which can often make the solution result converge to the global optimal solution better.

## 2.2 Model

In order to meet the practical needs of imaging through scattering media, a new encoder-decoder structure is designed. Through the skip connection, the local feature enhancement is further achieved, and the network structure is as shown in the figure below.

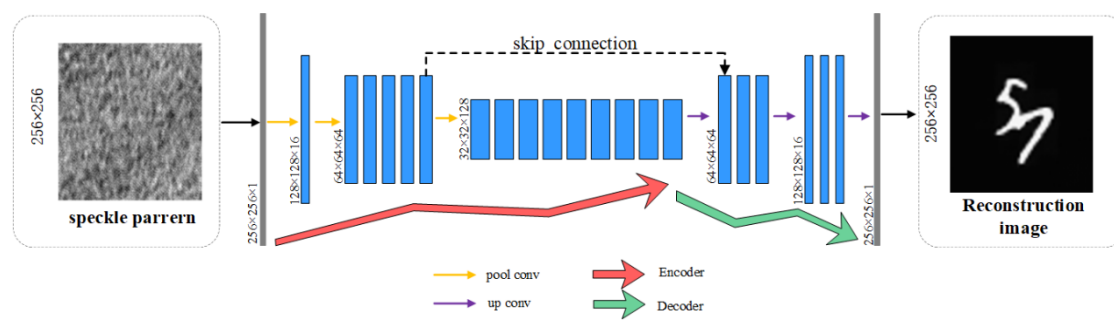

**Fig.1** The structure of the proposed PDSNet.

In solving the scattering problem in the range of OME, the commonly used network structure is U-shaped structure[10-12,15]. Its typical representative is U-net[16]. The network structure has a small number of parameters, which, on the one hand, is highly efficient in operation, and on the other hand, it also means that the network has limited reconstruction ability. Especially, when the complexity of solving the problem increases, it is obvious that the reconstruction effect decreases, as can be seen in the fig.4. At the same time, the increase in the number of parameters will also lead to a decrease in the computational efficiency, so the number of parameters is contradictory to the high computational efficiency. Here, we adopt the method of factorized convolution, and use n*1 and 1*n to replace the usual n*n convolution method. On the premise of ensuring the number of parameters, the computational complexity is reduced and the efficiency of network computing is improved.

The structure of encoder-decoder in U-shaped network is a classic point-to-point network architecture. However, high-level semantic information is mainly involved in the final prediction, so low-level information that fully represents local details cannot well participate in the final prediction process, resulting in insufficient detail of reconstruction results to a certain extent.

According to the principle of traditional iterative phase recovery algorithm, the initial image of the $k+1$ iteration is generated by constraining the output image of the last iteration pixel by pixel. Take HIO algorithm as an example:

$$\begin{cases} g_{k+1}(x,y) = g_k'(x,y), & for(x.y) \in \Gamma \\ g_{k+1}(x,y) = g_k(x,y) - \beta g_k'(x,y), & for(x.y) \notin \Gamma \end{cases} \quad (6)$$

Where, $k > 0$, $\Gamma$ is the set of points that satisfy the physical constraint, $g_k(x,y)$ is the input of iteration no. k, $g_k'(x,y)$ is the output of iteration no. k, $\beta$ is the feedback control parameter, which is similar to the learning rate in machine learning. Through several iterations, every pixel in the image is made satisfy the physical constraint as much as possible. The process of its reconstruction is the sequence from the local to the global.

Inspired by the above traditional algorithms, we hope that the underlying detailed information to be more involved in the final prediction process. We use the skip connection method to integrate low-level information with high-level information, so as to realize fully mining of pixel-level information and semantic-level features, and form a new convolutional neural network structure called PDSNet. The new encoder-decoder structure can integrate high-level semantic information with low-level local details, and improve the deficiency of the simple U-shaped structure in local details reconstruction.

## 3 Experiment

## 3.1 Simulation

Fig.1 shows the experiment setup in this paper. In the process of designing convolutional neural network, simulation data is used to test and verify the network performance.

The light signal of the target is modulated by the scattered medium after transmission in the first free space. After the second free space transmission process, the light signal emitted from the surface of the scattering medium reaches the camera, and the camera obtains the seemingly disordered light field intensity information. According to literature [17], The modulation effect of scattering medium on optical signal can be characterized by circular symmetric complex Gaussian random variables as characteristic matrix in the simulation process, of which variance is employed to represent the statistical characteristic of diffuser. Fresnel diffraction theory is applied to simulate two free space transmissions.

We hope to construct a scattering image reconstruction network that can break through the OME on the FOV constraint, therefore, it should be noted that all the simulation data and experimental data mentioned in this paper are beyond the range of FOV restricted by OME with target in grey-scale rather than binary. And the test environment of this paper is PyTorch 1.2.0 with RTX 2080Ti with I7-9700K CPU under ubuntu 16.04.

Mean absolute error (MAE), Structural Similarity Index Measure (SSIM) and Peak Signal to Noise Ratio (PSNR) are used as objective indicators to assess the effect of reconstruction in this paper. The definition of MAE is shown as follow:

$$MAE = \frac{1}{N}\sum_{i=1}^{N}|(f_i - y_i)| \qquad (7)$$

Where $f_i$ is the pixel value of the restored image, $y_i$ is the pixel value of the ground true (GT).

SSIM can be expressed in formula 8.

$$SSIM(x, y) = [l(x, y)]^{\alpha}[c(x, y)]^{\beta}[s(x, y)]^{\gamma} \qquad (8)$$

$$l(x, y) = \frac{2\mu_x\mu_y + c_1}{\mu_x^2 + \mu_y^2 + c_1} \qquad (9)$$

$$c(x, y) = \frac{2\sigma_{xy} + c_2}{\sigma_x^2 + \sigma_y^2 + c_2} \qquad (10)$$

$$s(x, y) = \frac{\sigma_{xy} + c_3}{\sigma_x\sigma_y + c_3} \qquad (11)$$

Where $l(x, y)$ is the brightness comparison, $c(x, y)$ is the contrast comparison, $s(x, y)$ is the structural comparison. $\mu_x$ and $\mu_y$ represent the mean value of x and y respectively, $\sigma_x$ and $\sigma_y$ represent the standard deviation of x and y respectively, $\sigma_{xy}$ is the covariance of x and y. $c_1$, $c_2$ and $c_3$ are constant to avoid a zero denominator.

$$PSNR = 10 \cdot \log_{10}(\frac{MAX_1^2}{MSE}) \qquad (12)$$

Where $MSE$ is the mean square error between the GT and the reconstructed image. $MAX_1$ is the maximum gray value that the image can reach.

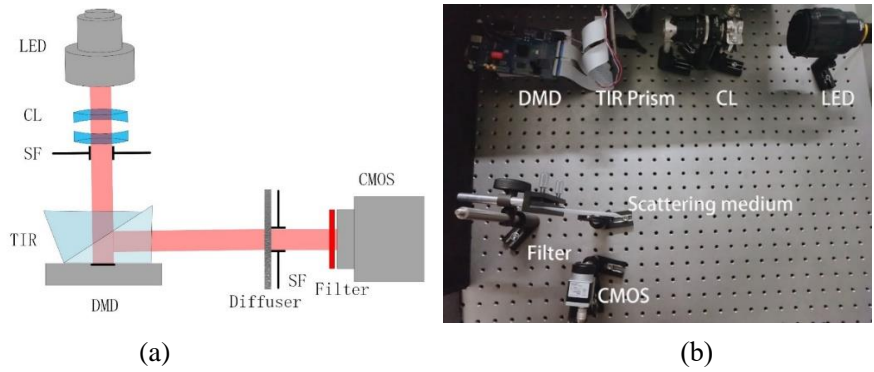

                    (a)                                 (b)

**Fig.2** Experiment setup uses an DMD as the object. (a) experiment setup; (b) the actual optical system.

To demonstrate PDSNet's ability to reconstruct targets under different conditions, five datasets are generated through simulation. The dataset 1-5 are named respectively as ***the no-overlap complex object dataset***, ***the overlap complex object dataset***, ***the media of same property dataset***, ***the media of different property dataset***, ***the human face dataset***. The cardinal targets commonly used in speckle correlation imaging based on deep learning are handwritten characters [10-16]. In order to improve the complexity of the targets to be restored in the first four datasets, the handwritten characters in MNIST are randomly combined in pairs. To be noticed, overlap strategy is utilized in several datasets to increase the diversity of targets as mentioned in table 1. The variances of characteristic matrixes constructed in ***the media of same property dataset*** are maintained the same value to simulate imaging with diffusers of the same property. A similar role is played by the variances of characteristic matrixes constructed in ***the media of different property dataset***. The targets in ***the human face dataset*** are derived from FEI Face Database [18]. The original targets acquired through the conditions shown in table 1 is utilized as the input of the optic system to generate the corresponding emulational speckle image, which constitutes the simulation datasets together.

The original targets of the datasets are employed as the GT of the neural network, while the speckle images are utilized as input images during the experiments. The size of both the GT image and the speckle image is 256*256.

**Table 1. The simulation datasets are generated according to the following conditions.**

|  | Dataset 1 | Dataset 2 | Dataset 3 | Dataset 4 | Dataset 5 |
|---|---|---|---|---|---|
| Basic element | MNIST | MNIST | MNIST | MNIST | FEI Face |
| With overlap characters | No | Yes | Yes | No | No |
| Number of diffusers | 1 | 1 | 10 | 10 | 1 |
| Variances of diffusers | One | One | Same | Different | One |
| Data pairs in training set | 3000 | 3000 | 30000 | 30000 | 350 |
| Data pairs in test set | 500 | 500 | 5000 | 5000 | 40 |

To demonstrate PDSNet's ability to reconstruct complex targets, PDSNet is tested with ***the no-overlap complex object dataset*** and ***the overlap complex object dataset***.

The relationship between autocorrelation of the original GT and autocorrelation of the speckle image does not meet formula 4 as shown in Fig.3(b) and (d), which proves that the target size in both datasets at this time is beyond the range of OME. Both datasets are respectively utilized to train PDSNet and U-net, and the reconstruction results of the test set are shown in Fig.3 (e) and Fig.3 (g) respectively. As shown in the following table, PDSNet obtained satisfactory results in both the two datasets, and it is better than U-net in both objective indicators and visual effects.

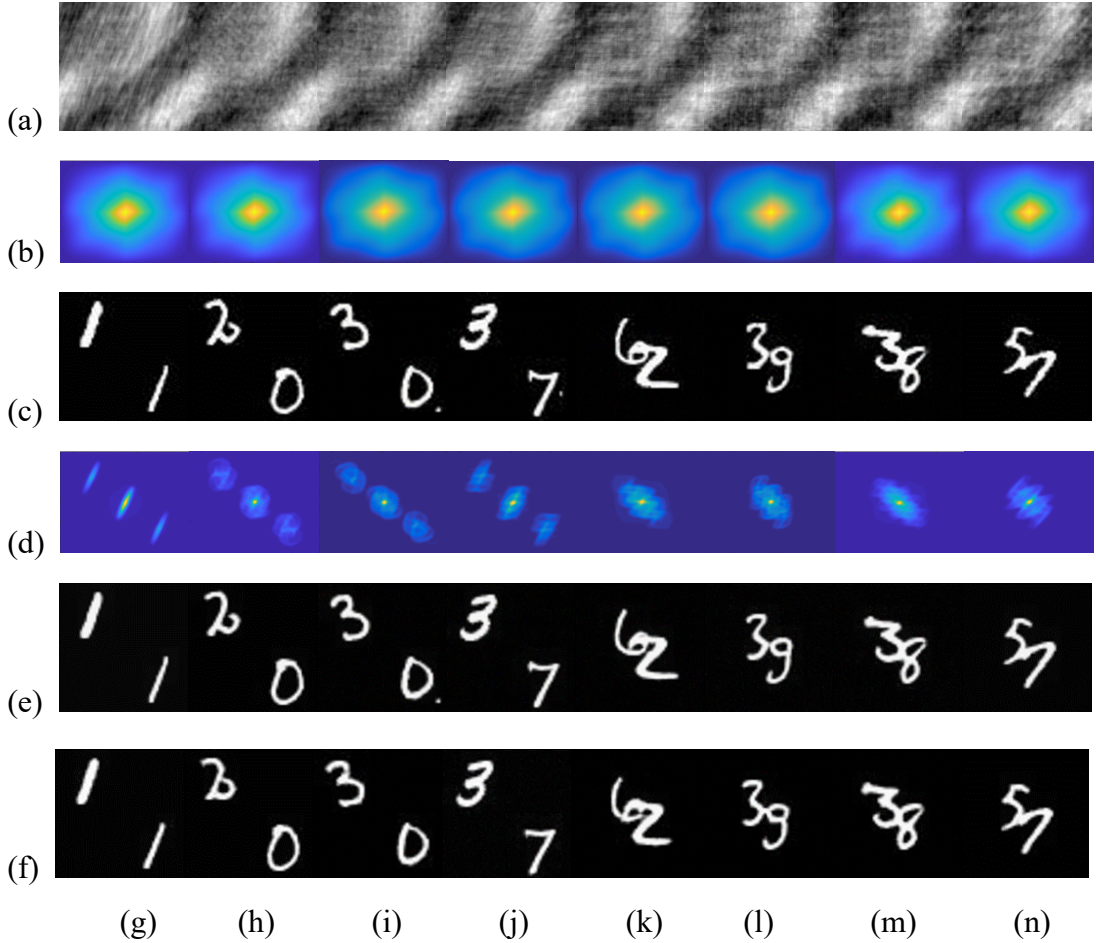

(a)

(b)

(c)

(d)

(e)

(f)

(g)　　(h)　　(i)　　(j)　　(k)　　(l)　　(m)　　(n)

**Fig.3** Test result of *the no-overlap complex object dataset* and *the overlap complex object dataset*: (a) speckle image; (b) autocorrelation of (a); (c)original target; (d) autocorrelation of (c); (e)output image of PDSNet; (f) output image of U-net; (g) ~ (j) belong to the experiment of *the no-overlap complex object dataset*; (k) ~ (n) belong to the experiment of *the overlap complex object dataset*.

Table 2. Average MAE, SSIM and PSNR of Two Complex Object Datasets.

|  |  | Objective Indicators | | |
|---|---|---|---|---|
|  |  | MAE | SSIM | PSNR |
| No-overlap complex object dataset | PDSNet | 0.0089 | 0.9400 | 39.9974dB |
|  | U-net | 0.0139 | 0.9191 | 37.4241dB |
| Overlap complex object dataset | PDSNet | 0.0102 | 0.9283 | 39.0627dB |
|  | U-net | 0.0126 | 0.9265 | 37.7578dB |

To further test the reconstruction capability of PDSNet for scattering media with the same statistical characteristics, PDSNet is tested with *the media of same property dataset*. The results of test set are shown in the Fig.4.

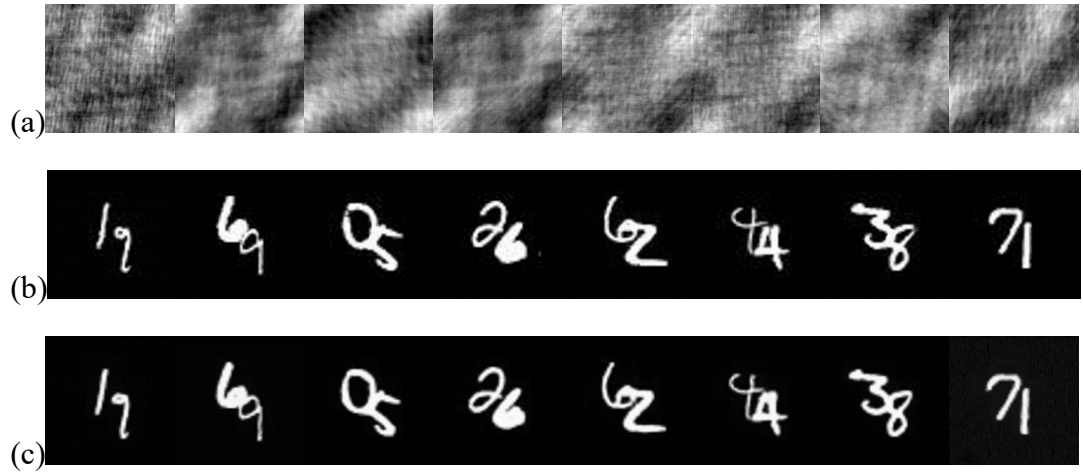

**Fig.4** Test result of *the media of same property dataset*: (a) speckle image; (b) original target; (c) output image of PDSNet.

Then, the reconstruction ability of PDSNet to scattering media with different statistical characteristics is tested, and *the media of different property dataset* is constructed for this purpose. The test results are shown in Fig.5.

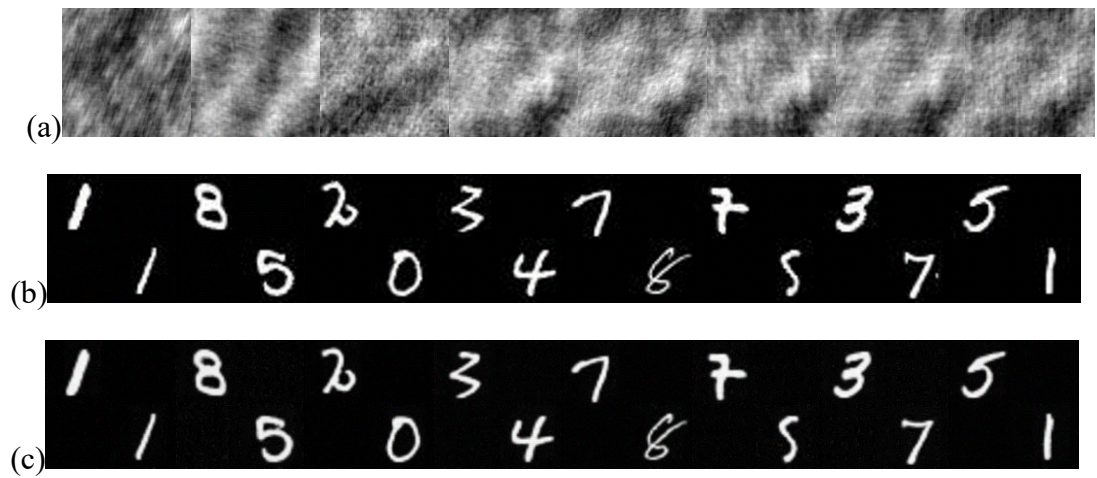

**Fig.5** Test result of *the media of different property dataset*: (a) speckle image; (b) original target; (c) output image of PDSNet.

**Table 3. Average MAE, SSIM and PSNR of Two Multi-media Datasets.**

|  | Objective Indicators | | |
| --- | --- | --- | --- |
|  | MAE | SSIM | PSNR |
| The same property media dataset | 0.0121 | 0.9235 | 37.9676dB |
| The different property media dataset | 0.0115 | 0.9152 | 38.9716dB |

The average MAE, SSIM and PSNR of *the media of same property dataset* and *the media of different property dataset* are shown in the table 3. For different scattering media, PDSNet has good target reconstruction ability beyond the range of OME.

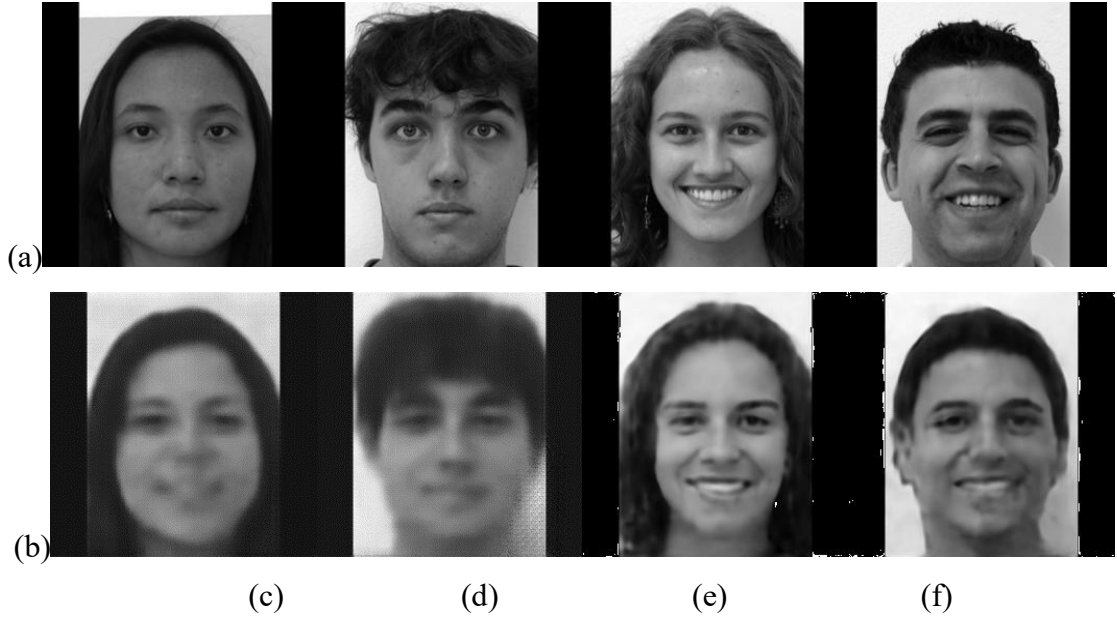

**Fig.6** Test result of *the human face dataset*: (a) original target; (b) output image; (c) and (d) results of PDSNet; (e) and (f) results of ResNet-50.

To further improve the complexity of the targets, *the human face dataset* is employed here. As shown in Fig.6(a) and (b), for the targets as detailed as a human face, PDSNet is not capable of reconstructing the image because of the limitation of its parameter amount. The same dataset is employed to test neural network with larger parameter amount using Resnet-50. As shown in Fig.6(c) and (d), better results are achieved by ResNet-50. Although PDSNet is sufficient to reconstruct handwritten character, larger parameter amount is to the benefit of restoring targets with more details like human faces shown by the contrast experiment between PDSNet and ResNet-50. As for the tasks with complex details, parameter amount of PDSNet can be enlarged by increasing numbers of layers.

## 3.2 Real System Experiment

In this section, data from the actual optical system is collected, and several trainings and tests are conducted.

The Fig.2(b) is the setup of the optical system. The light source is a LED with a central wavelength of 625nm (Thorlabs, M625L4). Using ground glasses (Thorlabs, DG100X100-220) as the scattering media, place it between the target to be measured and the camera plane. The final camera (Balser acA1920-155um) is responsible for collecting speckle images with information of the target to be restored. The distance between the target surface and the scattering medium is 25 cm, and the distance between the scattering medium and the camera is 8 cm. In the experiment, the target to be measured is generated by a DMD (resolution 1024*768 $pixels$, mirror element size 13.68 $\mu m / pixel$ ).

(1) module function analysis

In this section, the need for a skip connection is first verified. In order to demonstrate the necessity of the skip connection on the condition of targets size out of the range of OME, the handwritten characters in MNIST are randomly combined in pairs with overlap to generate 8000 images, which are utilized as the targets displayed on DMD. 7500 pairs data are contained in the training set while 500 pairs data are contained in the test set. It can be estimated from the Fig.7 that the autocorrelation distribution of the speckle images is quite different from that of the original GT image, which does not conform to formula 4, that is, the targets have exceeded the range of OME.

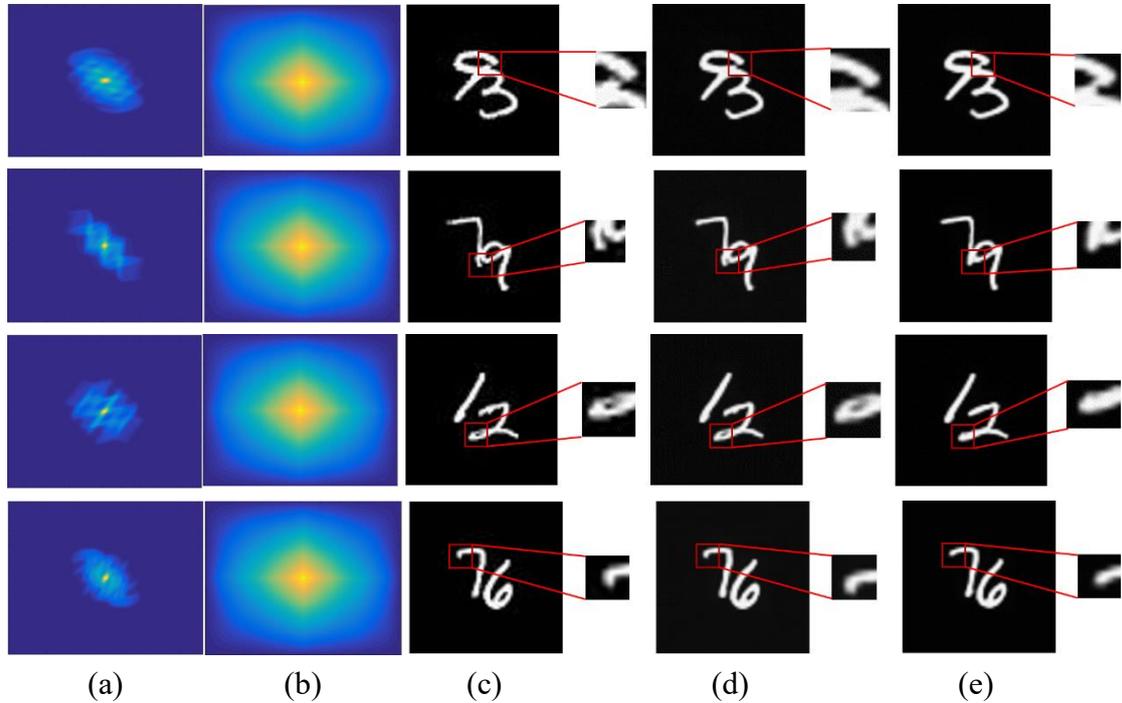

(a)　　　　(b)　　　　(c)　　　　(d)　　　　(e)

**Fig.7** Test result of PDSNet with and without skip connection: (a) autocorrelation of the original targets; (b) autocorrelation of the speckle images (c) original targets; (d) output images of PDSNet with skip connection; (e) output images of PDSNet without skip connection.

As can be seen from the comparison in Fig.7, the reconstructed results of PDSNet without skip connection basically restore the original target distribution. However, compared with GT, the detailed part of the target is not restored with great accuracy. It can be found that the reconstruction effect of the detailed part has been significantly improved relatively by skip connection.

In the downsampling process of encoder, PDSNet without skip connection gradually focuses on high-level semantic information, corresponding to the overall shape distribution of the target. This reconstruction strategy makes the reconstruction capability of PDSNet without skip connection limited. PDSNet without skip connection has some defects in the reconstruction ability of local details. The skip connections, which can introduce low-level details into the final prediction process, is of great necessity.

(2) reconstruct ability test

To test PDSNet's ability to reconstruct targets in untrained size, a *multi-scale dataset* is acquired. The experiment concludes that the range of OME is 150*150 pixels on DMD. Sizes of targets displayed on DMD are 300*300 pixels, 375*375 pixels, 420*420 pixels, 450*450 pixels, which all beyond the range of OME. Targets in size of 300*300 pixels, 375*375 pixels, 450*450 pixels are used as training set, when targets in size of 420*420 pixels are applied as test set keep untrained. Although the target size in the dataset is different, target in GT images are maintained the same size. 22500 data pairs are employed as training set, while 7500 data pairs are served as test set. As shown in Fig.8, PDSNet shows a very good generalization ability to reconstruct targets in untrained size.

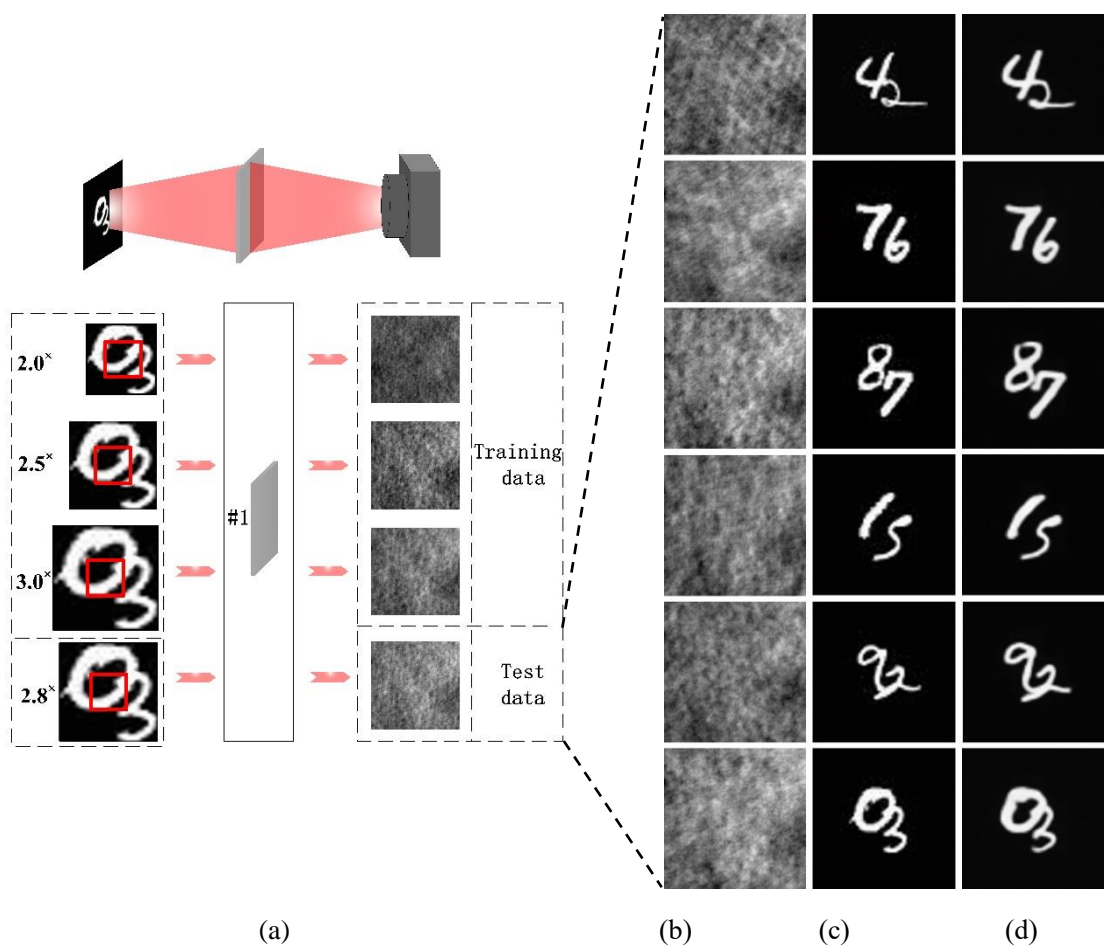

(a)　　　　　　　(b)　　(c)　　(d)

**Fig.8** Test PDSNet's ability to reconstruct targets in untrained size: (a) speckle patterns from different size of targets, red box shows the range of OME; (b)speckle images; (c) original targets; (d) output images of PDSNet.

To test PDSNet's ability to reconstruct targets through several scattering media, 4 ground glasses are employed as scattering media to collect speckle images for *multi-*

*diffusers dataset*. Targets used here are also acquired by combining two handwritten characters in MNIST with overlap. 30000 data pairs are employed as training set, while 2000 data pairs are used as test set. Test results and the original targets are shown in Fig.9.

The table 4 shows the average MAE, SSIM and PSNR of the test experiments carried out in this section. It can be seen from the table that PDSNet has a good reconstruct ability of untrained target scale. With single training, PDSNet can restore speckle images generated through several scattering media. The average FPS of PDSNet is 105, which is tested over 16000 images to be restored.

**Table 4. Average MAE, SSIM and PSNR of the Reconstruct Ability Experiments.**

|  | Objective Indicators | | |
| --- | --- | --- | --- |
|  | MAE | SSIM | PSNR |
| The ability to reconstruct untrained scale | 0.0207 | 0.8831 | 35.4188dB |
| The ability to reconstruct targets through different ground glasses | 0.0107 | 0.9249 | 38.6410dB |

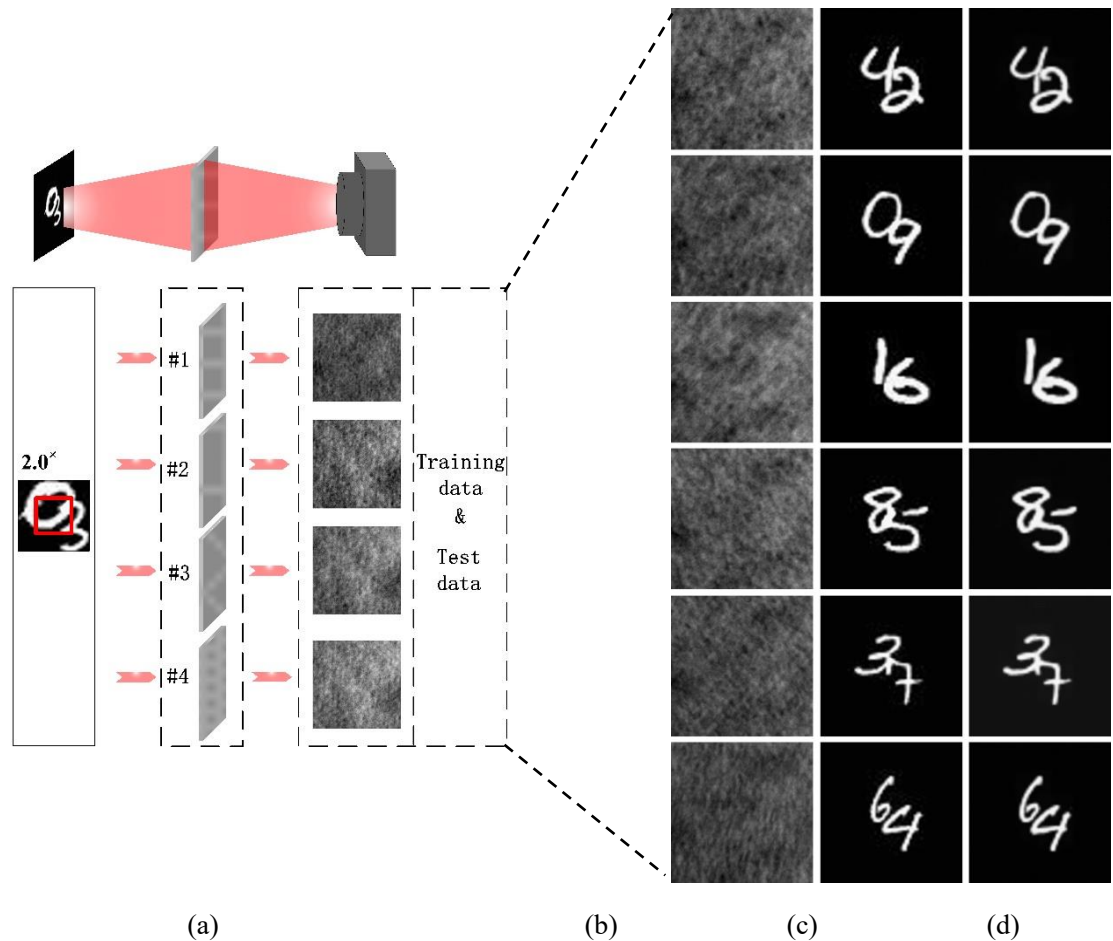

(a) (b) (c) (d)

**Fig.9** Test PDSNet's ability to reconstruct targets through several scattering media: (a) speckle patterns through different diffusers, red box shows the range of OME; (b) speckle images; (c) original targets; (d) output images of PDSNet.

# 4 Conclusion

Inspired by the traditional algorithm based on OME, this paper proposes a novel convolutional neural network called PDSNet for the reconstruction of scattered images. The experiments show that scattered images can be restored accurately by PDSNet in real-time, and good reconstruction ability is realized for the target with complex structure. Numbers of layers of PDSNet can be increased to improve amount of parameters to deal with targets with more details.

At the same time, PDSNet has a good generalization to reconstruct targets with untrained scales. For targets beyond the range of OME, the speckle images corresponding to targets at random scales can be reconstructed in one training, achieving "one-to-all" on the target scale. This ability enables us not to strictly require the target to be restored to have a uniform size, which is more conducive to the practical application of this technology in the future.

As a supervised learning method, PDSNet also needs to measure data in advance, but the good reconstruction ability of the network enables us to only measure scattering media more flexible in practical use. The test results of both simulation and experiment show that the proposed network also has good reconstruction potential for scattering media with different statistical characteristics. This is also a necessary ability for the practical application of deep learning to the reconstruction of scattering images of living biological tissues.

In the future, we will try to conduct experiments on living biological tissues that meet strong scattering conditions, hoping to be able to apply PDSNet to practical systems in fields such as medical imaging.